\documentclass[preprint]{vgtc}               % preprint
\ifpdf%                                % if we use pdflatex
  \pdfoutput=1\relax                   % create PDFs from pdfLaTeX
  \usepackage{graphicx}                % allow us to embed graphics files
  \DeclareGraphicsExtensions{.pdf,.png,.jpg,.jpeg} % for pdflatex we expect .pdf, .png, or .jpg files
\else%                                 % else we use pure latex
  \usepackage{graphicx}                % allow us to embed graphics files
  \DeclareGraphicsExtensions{.eps}     % for pure latex we expect eps files
\fi%

\graphicspath{{figures/}{pictures/}{images/}{./}} % where to search for the images

\usepackage{microtype}                 % use micro-typography (slightly more compact, better to read)
\PassOptionsToPackage{warn}{textcomp}  % to address font issues with \textrightarrow
\usepackage{textcomp}                  % use better special symbols
\usepackage{mathptmx}                  % use matching math font
\usepackage{times}                     % we use Times as the main font
\usepackage{cite}                      % needed to automatically sort the references
\usepackage{tabu}                      % only used for the table example
\usepackage{booktabs}                  % only used for the table example
\usepackage{enumitem}
\onlineid{1120}

\vgtccategory{Research}

\vgtcinsertpkg

\preprinttext{To appear in proceedings of the 29th IEEE Conference on Virtual Reality and 3D User Interfaces Abstracts and Workshops 2022 (VRW)}

\title{Comparing Controller With the Hand Gestures Pinch and Grab for Picking Up and Placing Virtual Objects}

\author{Alexander Schäfer\thanks{e-mail: alexander.schaefer@dfki.de}\\ %
        \scriptsize TU Kaiserslautern %
\and Gerd Reis\thanks{e-mail: gerd.reis@dfki.de}\\ %
     \scriptsize German Research Center for Artificial Intelligence %
\and Didier Stricker\thanks{e-mail: didier.stricker@dfki.de}\\ %
     \parbox{1.4in}{\scriptsize \centering German Research Center for Artificial Intelligence \\ TU Kaiserslautern}}

\teaser{
  \includegraphics[width=1.0\textwidth]{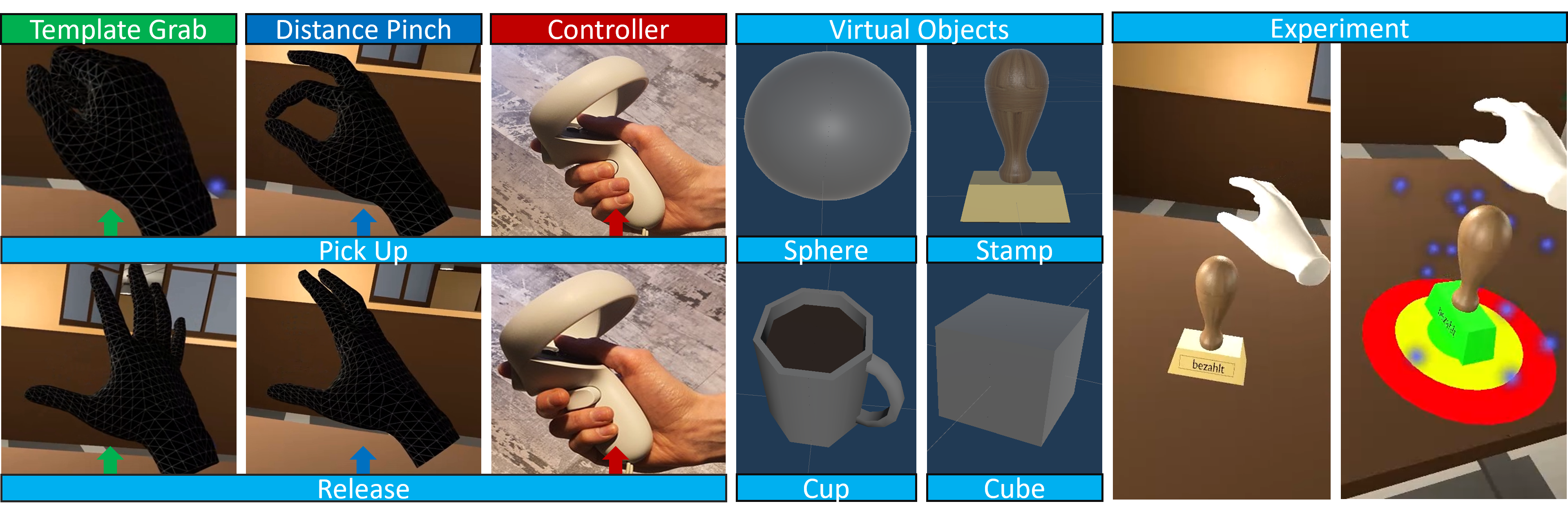}
  \caption{The different techniques for picking up as well as the virtual objects used in the experiment.}
  \label{fig:Teaser}
}

\abstract{Grabbing virtual objects is one of the essential tasks for Augmented, Virtual, and Mixed Reality applications. Modern applications usually use a simple pinch gesture for grabbing and moving objects. However, picking up objects by pinching has disadvantages. It can be an unnatural gesture to pick up objects and prevents the implementation of other gestures which would be performed with thumb and index. Therefore it is not the optimal choice for many applications. In this work, different implementations for grabbing and placing virtual objects are proposed and compared. Performance and accuracy of the proposed techniques are measured and compared.

} % end of abstract

\CCScatlist{
  \CCScatTwelve{Human-centered computing}{Human computer interaction (HCI)}{Interaction techniques}{Gestural input};
  \CCScatTwelve{Human-centered computing}{Human computer interaction (HCI)}{Interaction paradigms}{Mixed / Augmented Reality; Virtual Reality};
}

\begin{document}


\firstsection{Introduction}

\maketitle

%% \section{Introduction} %for journal use above \firstsection{..} instead
Modern Head Mounted Displays (HMD) for Augmented (AR), Virtual (VR), and Mixed Reality (MR) applications support hand tracking out of the box. Manufacturers provide a small set of gestures for developers to implement interactions. The pinch gesture is the most common gesture for grasping and interacting with virtual objects. 
%The most eminent examples of this are the Mixed Reality Toolkit \cite{mrtk}, Oculus SDK \cite{oculusSDK}, and the Leap Motion SDK \cite{leapSDK}.  
Pinching can be easily implemented by utilizing the distance of thumb and index finger which is reliable even with noisy hand tracking data. Furthermore, it is a gesture which can be easily performed by users. However, pinching with two fingers is not easily distinguishable from other gestures. Each gesture that requires thumb and index to be close to each other will result in a recognized pinch gesture. For example, an OK gesture would automatically trigger a pinch gesture.  Hand gestures can also be implemented by using a template based approach which can store and compare a hand shape. This approach allows developers more freedom for specific gestures such as an OK gesture that is not recognised as a pinch. However, it is assumed that a pinch gesture is easier to perform and therefore more accurate and efficient. Therefore, this paper investigates whether a template-based hand gesture is an adequate alternative for pinching when grabbing objects. 
%In particular, it will be investigated whether a template matching based gesture can compete with a finger distance based gesture.  
%\par 
%This work compares the pinch and a grab gesture for picking up virtual objects in AR,VR, and MR applications. 
Picking up objects with a controller was included for comparison. The three techniques will be compared in terms of accuracy and efficiency: Distance Pinch, Template Grab, and Controller (Depicted in Figure \ref{fig:Teaser}).

\section{Implementation}
\label{sec:implementation}
The implementation for each technique is briefly described. 
%It is to note that the pick up range was 20 cm within controller or hands. Objects were highlighted once they were in range for pick up. 
\par 
\noindent \textbf{Distance Pinch.}
The Distance Pinch gesture was implemented by utilizing thumb and index tip positions provided from the chosen hand tracking solution. The distance between those two points is measured for each frame. A threshhold determines if a user is currently pinching. 
%Three states can be distinguished: pinching, not pinching, and holding. Where holding means a pinch has been performed and the distance between thumb and index finger has not gone above the threshhold. 
If thumb and index finger are closer than 3 cm to each other, pinching is activated (this is an empirical value). It was found that there was a critical area where noise affected recognition. For example, if the fingers were held about 3 cm apart (2.9 - 3.1 cm), the system would jump between pinch detected and no pinch detected. Therefore, it was introduced that the state of the gesture would only change if the same value was reported for 100 ms which resulted in an overall much smoother user experience. \par 

\noindent \textbf{Template Grab.}
The Template Grab gesture was implemented by utilizing all hand joints provided by the hand tracker (23 points). The distance between hand joints is stored in order to recognise specific hand shapes. This allows recognition for static gestures by comparing frames from the hand tracker with the stored distances. Two different static gestures are required for this approach: One for initiating the grab and one for releasing the object. Therefore, a static gesture resembling a closed hand and a static gesture with a relaxed hand are stored. The gestures are rotation invariant. Detecting a closed hand will initiate a grab event to nearby objects while detecting a relaxed hand will release the currently grabbed object. It was decided to include two static gestures for the release state: One with the hand partially open and one with the hand fully open. Ideally, the gesture with the hand partially open releases the object.\par

\noindent \textbf{Controller.}
Grabbing and releasing an object with the controller is performed by pressing the grip button on the VR controller. 

\section{Evaluation}
\label{sec:eval}
\noindent \textbf{Apparatus.} Hand tracking and controllers were provided by the Oculus Quest 2 VR HMD. \par

\noindent \textbf{Objectives.} It should be investigated how Distance Pinch compares to Template Grab. Quantitative metrics to measure performance and accuracy are employed. Performance is measured by the time required from grabbing to placing an object. The accuracy is measured by how close object was placed to the center of the target. \par
\noindent \textbf{Participants.} 18 participants (6 female) participated in the experiment (Age $\mu$ = 33.5) and their self-assessed experience with VR is $\mu$ = 2.1 (Answered with 5-Point-Likert Scale, higher value means higher experience). 
%The study was conducted according to the guidelines of the Declaration of Helsinki. Due to the COVID-19 pandemic, only a limited amount of participants could be recruited. The experiment was conducted under strict hygiene procedures, including multiple VR-HMDs and disinfection of the equipment after each experiment.\par

\noindent \textbf{Experimental Task.}
The experiment is a within-subjects design. Participants need to grab and subsequently place virtual objects on specific positions on a table. The objects must be placed on a target consisting of three parts: green, yellow and red. The colours are arranged from the inside of the target to the outside (See Figure \ref{fig:Teaser} right side). The participants were told to try to place the objects as fast and as close to the center of the target as possible. Depending on how close the object is placed to the centre of the target, the colour of the object changes accordingly. This gives the subject the opportunity to understand how well the object has been placed.  A total of four objects must be placed 10 times each. This is repeated for each grab technique. The virtual objects are: Cube, Sphere, Cup, and Stamp. The experiment was conducted in seating position. \par % and required the participants to stretch their arms in order to place the objects. \par
\noindent \textbf{Procedure.}
%Informed consent was handed to the participants before the experiment. 
The order of pick up techniques was counterbalanced using the Balanced Latin Square algorithm. Subjects were allowed to try each method for a short time before the actual task began (as many had neither VR nor hand tracking experience). After a total of 40 objects were placed with each technique, resulting in 120 objects placed by a participant, the experiment concluded.
\section{Results and Discussion}
\label{sec:results}
\begin{figure}[t]
\centering
\includegraphics[width=0.5\textwidth]{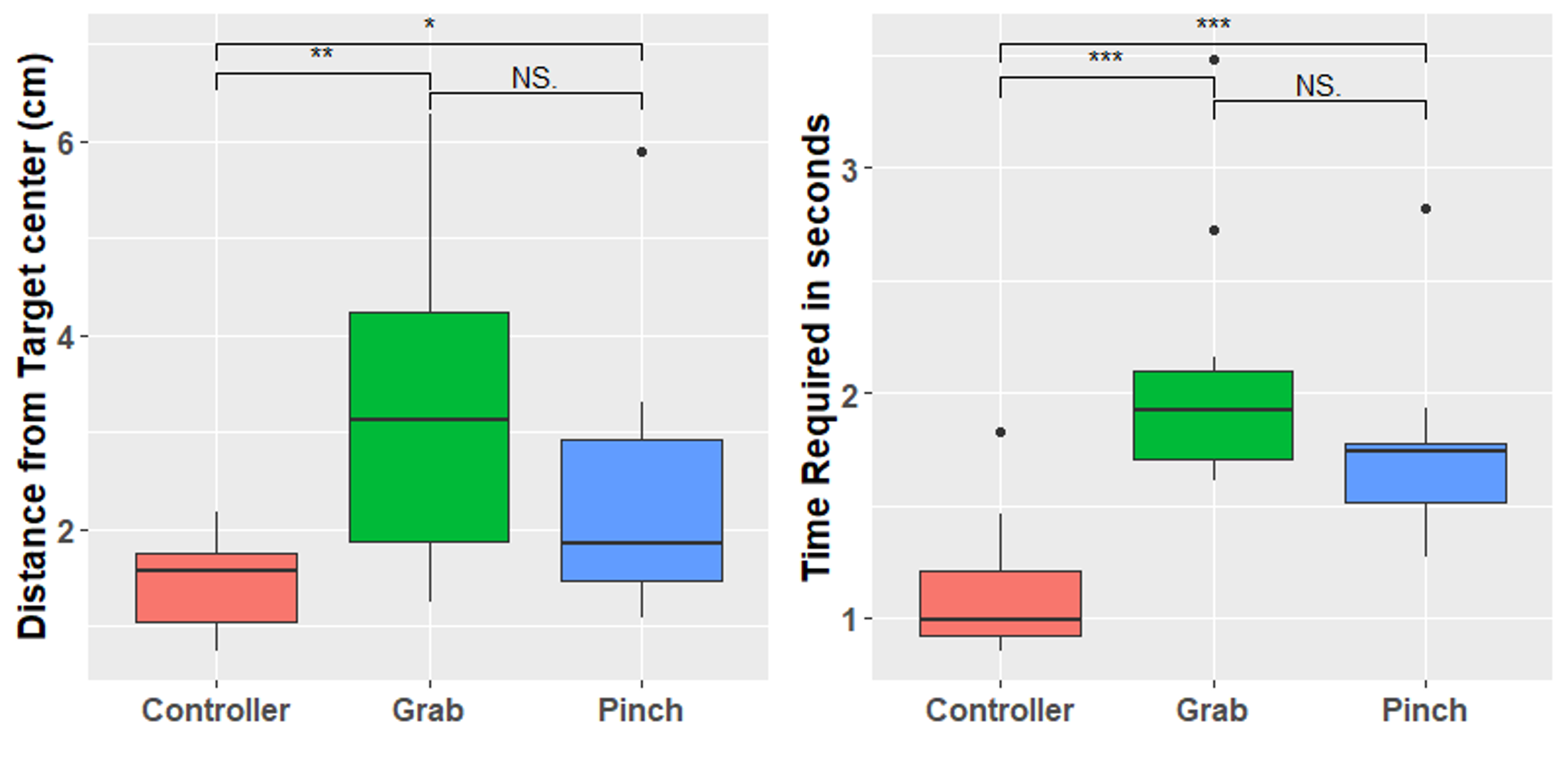}
\caption{Time required by participants from grabbing to placing the objects (left) and distance placed to the center of the target (right). Significance Levels: *** = 0.001; ** = 0.01; * = 0.05; NS = No Significance}
\label{fig:Dataplot}
\end{figure}

A total of 2.160 virtual objects were picked up and placed for the study (720 per technique). The averaged results of the dependent variables \textit{Accuracy} and \textit{Task Completion Time} are shown in \ref{fig:Dataplot}. Levene's test assured the homogeneity of the input data ($p > 0.05$) and therefore one-way ANOVA was used for statistical analysis. The ANOVA result ($F(2,51) = 16.95 ; p < 0.001$) showed significant differences between the techniques. Tukey's Honest Significant Difference (TukeyHSD) was used as post hoc analysis of the data. Significance between techniques is depicted in Figure \ref{fig:Dataplot}. Picking up and subsequently placing an object was significantly faster with Controller as compared to the other techniques. Controller was also significantly more accurate than Grab and Pinch. Grab and Pinch are not significantly faster or more accurate compared to each other.
\setlength{\belowcaptionskip}{-15pt}

\section{Conclusion and Key Takeaways}
The results show that picking up and subsequently placing objects with a Controller performed significantly better compared to using bare hands with the proposed techniques. This lines up with results from other studies comparing interactions with controller and hand gestures as input modalities \cite{masurovsky2020controller,hameed2021evaluating}. However, no significant differences were found between Distance Pinch and Template Grab regarding performance and accuracy. \par
By analysing the results, interviewing and observing the participants, the following takeaways can be formulated in order to help researchers and developers while implementing virtual object interactions:
\begin{itemize}[leftmargin=*,topsep=2pt, itemsep=1pt]
    \item Participants tried to grab an object (by performing the gesture or pressing the grip button) before being close enough to the object. This resulted in not picking up the object (this happened multiple times for many participants). It is to note that objects were highlighted once they were in range for picking up. The objects could be picked up if they were within 20 cm of the hand or controller.
    \item Participants had the impression that the objects were "glued" to their hand when using hand gestures. For the pinching gesture this meant that the thumb and index distance threshold had not yet been exceeded when trying to place. For the template gesture, participants often did not open their hand all at once, but finger by finger. This "glue" effect was stronger with the Template Grab resulting in rather poor accuracy, which can also be observed in Figure \ref{fig:Dataplot}.
    \item The shape of an object should be considered when designing specific gestures for picking up objects. For example, almost all participants tried to grab the cup by the handle and the stamp by the knob. Pinch would be better for the cup and Grab better for the stamp but neither the Pinch nor the Grab gesture is suited for picking up both objects intuitively.
    %\item It was observed that adding the controller as a baseline had a negative impact on the ratings for subsequent techniques. 
\end{itemize}
Static gestures should be further explored for picking up objects. By employing gestures with template matching, gestures specifically tailored to picking objects at certain points could be implemented. For example, a Pinch when trying to grab the handle of a cup and Grab when trying to pick up the cup from the top. This work aims towards natural and intuitively grabbing virtual objects.
\acknowledgments{
Part of this work was funded by the Bundesministerium für Bildung und Forschung (BMBF) in the context of ODPfalz under Grant 03IHS075B. This work was also supported by the EU Research and Innovation programme Horizon 2020 (project INFINITY) under the grant agreement ID: 883293.}

\bibliographystyle{abbrv-doi}

\bibliography{template}

\begin{thebibliography}{1}

\bibitem{hameed2021evaluating}
A.~Hameed, A.~Perkis, and S.~Möller.
\newblock Evaluating hand-tracking interaction for performing motor-tasks in vr
  learning environments.
\newblock In {\em 2021 13th International Conference on Quality of Multimedia
  Experience (QoMEX)}, pp. 219--224, 2021. doi: {{%
10\hspace{.1pt}\discretionary{.}{%
}{.}\hspace{.4pt}1109\discretionary{/}{%
}{/}QoMEX51781\hspace{.1pt}\discretionary{.}{%
}{.}\hspace{.4pt}2021\hspace{.1pt}\discretionary{.}{%
}{.}\hspace{.4pt}9465407}}


\bibitem{masurovsky2020controller}
A.~Masurovsky, P.~Chojecki, D.~Runde, M.~Lafci, D.~Przewozny, and M.~Gaebler.
\newblock Controller-free hand tracking for grab-and-place tasks in immersive
  virtual reality: Design elements and their empirical study.
\newblock {\em Multimodal Technologies and Interaction}, 4(4), 2020. doi: {{%
10\hspace{.1pt}\discretionary{.}{%
}{.}\hspace{.4pt}3390\discretionary{/}{%
}{/}mti4040091}}


\end{thebibliography}
\end{document}